\documentclass[10pt,letterpaper]{article}
\usepackage[top=0.85in,left=2.75in,footskip=0.75in,marginparwidth=2in]{geometry}

\usepackage[utf8]{inputenc}

\usepackage{cite}

\usepackage{nameref,hyperref}

\usepackage{microtype}
\DisableLigatures[f]{encoding = *, family = * }

\raggedright
\usepackage{changepage}

\usepackage[aboveskip=1pt,labelfont=bf,labelsep=period,singlelinecheck=off]{caption}

\makeatletter
\renewcommand{\@biblabel}[1]{\quad#1.}
\makeatother

\usepackage{lastpage,fancyhdr,graphicx}
\usepackage{epstopdf}
\fancyheadoffset[L]{2.25in}
\fancyfootoffset[L]{2.25in}

\usepackage{color}

\definecolor{Gray}{gray}{.25}

\usepackage{graphicx}

\usepackage{amsmath}
\usepackage{amssymb}
\usepackage{latexsym}
\usepackage{amsthm}
\usepackage{hyperref}

\newcommand{\onlinecite}[1]{[\hspace{-1ex} \citenum{#1}]}

\newcommand{\editor}[2]{%
  \expandafter\newcommand\csname #1note\endcsname[1]{%
    \textcolor{#2}{\textbf{#1:} ##1}}%
  \expandafter\newcommand\csname #1\endcsname[1]{%
    \textcolor{#2}{##1}}%
  \expandafter\newcommand\csname #1cancel\endcsname[1]{%
    \textcolor{#2}{\sout{##1}}}%
  \expandafter\newcommand\csname #1change\endcsname[2]{%
    \textcolor{#2}{\sout{##1} ##2}}%
  \newenvironment{#1text}{\color{#2}}{\color{black}}
}

\usepackage{sidecap}

\usepackage{wrapfig}
\usepackage[pscoord]{eso-pic}
\usepackage[fulladjust]{marginnote}
\reversemarginpar

\begin{document}
\vspace*{0.35in}

\begin{flushleft}
{\Large
\textbf\newline{ Sampling molecular conformers in solution with quantum mechanical
  accuracy at a nearly molecular mechanics cost}
}
\newline
\\
Marta Rosa\textsuperscript{1},
Marco Micciarelli\textsuperscript{1},
Alessandro Laio\textsuperscript{1},
Stefano Baroni\textsuperscript{1,*}
\\
\bigskip
\bf{1} SISSA -- Scuola Internazionale Superiore di Studi
  Avanzati \\ via Bonomea 265, 34136 Trieste -- Italy
\\
\bigskip
* baroni@sissa.it

\end{flushleft}

\begin{abstract}
  We introduce a method to evaluate the relative populations of
  different conformers of molecular species in solution, aiming at
  quantum mechanical accuracy, while keeping the computational cost at
  a nearly molecular-mechanics level. This goal is achieved by
  combining long classical molecular-dynamics simulations to sample
  the free-energy landscape of the system, advanced clustering
  techniques to identify the most relevant conformers, and
  thermodynamic perturbation theory to correct the resulting
  populations, using quantum-mechanical energies from
  density-functional theory.  A quantitative criterion for assessing
  the accuracy thus achieved is proposed. The resulting methodology is
  demonstrated in the specific case of cyanin
  (cyanidin-3-glucoside) in water solution.
\end{abstract}

\newpage
\section{Introduction} \label{sec:introduction} 

Organic molecules in water solution usually exist in several conformations, separated from each other by high free-energy barriers. Determining the relative population of these conformers is key for predicting molecular properties such as, \emph{e.g.}, optical or NMR spectra. The procedure that is normally followed is to estimate these populations starting from the relative energies of the conformers, without treating explicitly the solvent molecules. The effect of the solvent can then be accounted for by using an implicit solvation scheme \cite{tomasi2005quantum, mennucci1997continuum, dupont2013self, barone1997new, andreussi2012revised, orozco2000theoretical} while entropic effects can be estimated in the harmonic approximation from the vibrational frequencies of the solute \cite{andricioaei2001calculation, carlsson2006calculations, simonson2002free, suarez2015direct}. This procedure is computationally expedient and provides an estimate of the populations that can at times be rather accurate. However, in many cases the procedure is affected by large systematic errors, due to the intrinsically molecular nature of the solvent. For example, a specific conformer can be stabilized by the presence of a solvent molecule bridging two moieties of the solute, a situation that would be missed by any implicit-solvent scheme, or the presence of floppy vibrational modes could make the use of the harmonic approximation questionable.

In this paper we propose an approach that allows estimating the relative populations of various conformers with the accuracy of ab initio (AI) molecular dynamics (MD) in explicit solvent at the cost of a few thousand quantum calculations. The configurational space of the solvated molecule is sampled by long molecular-mechanics (MM) MD runs, while density functional theory (DFT)  calculations are performed only on a carefully selected set of configurations. In our approach a recently proposed clustering algorithm \cite{rodriguez2014clustering} is first applied to a long, supposedly ergodic, MMMD trajectory, to identify molecular conformers corresponding to suitably defined slow collective variables, and the relative populations are then estimated from the resulting residence times. In order to estimate quantum mechanical (QM) corrections to the population of these conformers, we exploit first-order thermodynamic perturbation theory, a procedure first pioneered by Warshel \cite{muller1995ab, wesolowski1994ab}, using the QM energies computed at the DFT level on a set of uncorrelated configurations for each conformer. A key prerequisite for the success of this procedure is a high level of consistency between the MM and the QM free-energy landscapes. In particular, the stable conformers must be structurally similar at the MM and QM levels, while their populations can differ substantially. Importantly, we show that the magnitude of the second-order corrections to the conformational free energies, computed without taking into account the solvent, while not used directly to evaluate populations, provides a fair criterion to appraise \emph{ex-post} the reliability of first-order corrections. Our approach is demonstrated by applying it to the cyanin (cyanidin-3-glucoside) molecule in water solution at room temperature.

\section{Methodology} \label{sec:methodology} 

The configurational space of large molecules in solution is made of several free-energy basins, (\emph{molecular conformers}) separated by high free-energy barriers. These systems can be easily described with MM simulations, and the relative populations of different conformers estimated from the residence times of the molecule in each of them. Nevertheless, quantum accuracy in the inter-atomic forces is often needed, and sampling the configuration space with AIMD is hindered by the long time needed for the molecule to hop between any two conformers.

Here we describe an approach that allows one to estimate the relative populations of different conformers and to determine statistical averages of various observables at the QM level from MMMD trajectories. In the particular case where one is just interested in the relative populations of different conformers, the following derivation will lead to a direct estimate of the relevant free energy differences (see Eqs. \ref{eq:boltz} and \ref{eq:sviluppo}).

In classical statistical mechanics, the physical properties of a system can be expressed as time averages of suitably defined configuration-space functions (\emph{observables}) over molecular trajectories:

\begin{align}
  \langle A\rangle &= \lim_{\tau \rightarrow \infty} 
                     \frac{1}{\tau} \int_0^\tau
                     A(q_t)dt, \label{eq:<A>micro} \\  
                   & = \frac{1}{Z} \int A(q)
                     \mathrm{e}^{-\frac{V(q)}{k_B T}}
                     dq, \label{eq:<A>canon}
\end{align}

where $q$ indicates the set of atomic coordinates, $V(q)$ is the potential energy function, $\tau$ the length of the trajectory, $T$ the system's temperature, $k_B$ the Boltzmann's constant, and $Z$ the partition function. Molecular conformers in configurational space are defined in such a way that thermalization within each of them occurs in typical molecular-vibration times, whereas the hopping between any two of them is kinetically hindered by free energy barriers: the corresponding transition rate follows therefore an Arrhenius law, typical of thermally activated processes. As a consequence, global thermalization requires simulation times that are exponentially long in the height of the free-energy barriers, and can thus be hardly achieved using accurate QM methods, whose scope is limited to processes spanning hundreds of pico-seconds at most. Methods based on MMMD extend by several orders of magnitude the time scales accessible to molecular simulations, bringing them up to the micro-second range and up, but their accuracy is limited by the quality of the empirical force fields employed therein.

In those cases where classical force fields are accurate enough to describe the general topography of the free-energy landscape, the computational convenience of MMMD can actually be combined with the accuracy of AIMD, by means of thermodynamic perturbation theory. In order to proceed, let us start from Eq. \eqref{eq:<A>canon} and rewrite it as:
\begin{align}
  \langle A\rangle &\approx \sum_\mathcal{C} \frac{Z^\mathcal{C}}{Z} \frac{1}{Z^\mathcal{C}}
                     \int_{q\in \mathcal{C}} A(q)
                     \mathrm{e}^{-\frac{V(q)}{k_B T}}
                     dq, \label{eq:<A>conf} \\
                   &\approx \sum_\mathcal{C} p^\mathcal{C}
                     \frac{1}{\tau^\mathcal{C}}\int_0^{\tau^\mathcal{C}}
                     A(q_t)dt, \label{eq:<A>multi} 
\end{align} where the sums extend over all the conformers $\mathcal{C}$; in Eq. \eqref{eq:<A>conf} the integrals are restricted to the portions of configurational space characterizing a given conformer and $Z^\mathcal{C}$ is the corresponding restricted partition function; in Eq. \eqref{eq:<A>multi} $p^\mathcal{C} = \frac{Z^\mathcal{C}}{Z} \approx \frac{\tau^\mathcal{C}}{\tau}$ is the probability that the system is found in conformer $\mathcal{C}$ and the time averages are evaluated over times $\tau^\mathcal{C}$ longer than the local thermalization time, but significantly shorter than the typical residence time within that same conformer. These restricted averages are accessible to AIMD, whereas evaluating $p^\mathcal{C}$ would require long trajectories that can only be generated via MMMD. In order to overcome this difficulty, we express these probabilities in terms of the conformers' free energies, $F^\mathcal{C}$, defined as:
\begin{equation}
  p^\mathcal{C}\propto \mathrm{e}^{-\frac{F^\mathcal{C}}{k_B T}}. \label{eq:Fdef}
\end{equation}
Eq. \eqref{eq:Fdef} allows us to express the ratio between the populations of a same conformer computed at different levels of theory in terms of the exponential of the corresponding free-energy differences,\cite{zwanzig1954high} such as resulting from two different MM force fields or a force field and a QM approach, such as DFT. In the latter case, with obvious notation one would have:
\begin{equation}
  F^\mathcal{C}_{QM}= F^\mathcal{C}_{MM} +k_BT \: \log \left\langle
  \mathrm{e}^{\frac{1}{k_BT} (E_{QM} - E_{MM})}
  \right\rangle^\mathcal{C}_{MM}, \label{eq:boltz} 
\end{equation}
where $\langle\cdot\rangle^\mathcal{C}_{MM}$ indicates a time (or canonical) average performed within the $\mathcal{C}$ conformer at the MM level. Eq. \eqref{eq:boltz} is in principle exact, if the $\mathcal{C}$ conformer is well defined both at the MM and QM levels, which obviously implies that the two levels of theory are close enough. Even in this case the evaluation of the statistical average of the exponential in Eq. \eqref{eq:boltz} is severely hampered by thermal fluctuations in the exponent, which scale as the system size when solvent molecules are explicitly accounted for (more on reducing the impact of solvent energy fluctuations in the following). What can be done in practice is expanding the logarithm in Eq. \eqref{eq:boltz} in a series of cumulants, in the spirit of perturbation theory, and retaining only the linear term, which converges relatively easily in water:

\begin{equation}
  F^\mathcal{C}_{QM} \approx F^\mathcal{C}_{MM} + 
  \langle   \Delta E \rangle 
  +\mathcal{O}\left (\Delta E^2 \right ),
  \label{eq:sviluppo}
\end{equation}

where $\Delta E= E_{QM} -E_{MM}$ is the difference between the QM and MM energies.

The second order correction to Eq. \eqref{eq:sviluppo} is given by $\frac{\kappa_2}{2 k_B T}$, where $\kappa_2 =  \langle   \Delta E^2  \rangle -  \langle   \Delta E \rangle^2 $ is the second cumulant of $ \Delta E  $ and all the averages are sampled in the $\mathcal{C}$ conformer on the MM distribution of molecular configurations. 

Second and higher-order cumulants converge worse and worse as their order increases. Already at second order, the energy fluctuations due to the solvent are too large to compute the second cumulant for systems comprising a few hundred solvent molecules. An option could be to account for the effects of a few solvent molecules that interact most strongly with (\emph{i.e.} that are closest to) the solute. Beside the intrinsic arbitrariness of this procedure, its most important drawback would be the difficulty to estimate its accuracy. We decided therefore to stick to first order in Eq. \eqref{eq:sviluppo} to correct the relative populations of different conformers, while second cumulants, as computed neglecting solvent effects (\emph{i.e. in vacuo}), are used to estimate the accuracy of this first-order approximation, using the procedure outlined below.

The contribution of the second cumulants to the relative populations vanishes if their value is the same for the different conformers, as the relative free energies would not be affected by them in this case.
The relative magnitude of the second cumulants for different conformers can however be reliably estimated by computing them for \emph{dehydrated} molecular configurations, \emph{i.e.} by comparing QM and MM energies corresponding to the molecular configurations generated by an explicit-solvent MMMD simulation, upon stripping off solvent molecules.\footnote{Second cumulants can also be reliably estimated in a QM/MM scheme, because in this case energy fluctuations due to the solvent can be expediently removed relying on correlated sampling.} The cumulants thus obtained obviously cannot be used to improve our estimate of QM corrections to the relative populations of different conformers, but they do provide a fair estimate of the accuracy of the first-order approximation to this correction. A further piece of information that comes from the calculation of second cumulants is their convergence rate, \emph{i.e.}  their variance within a same conformer: when the MM and QM free-energy landscapes are similar, the value of the second cumulant of a conformer will easily converge, at least when neglecting solvent effects, while this will be increasingly difficult as the MM and QM landscapes differ from each other. All in all, second cumulants determine the accuracy of the first-order corrections to the relative populations in three distinct ways: \emph{i)} The statistical error of the first cumulant within a same conformer is given by the second cumulant divided by the number of independent molecular configurations generated in the MMMD run, $\Delta F^\mathcal{C}_1 = \sqrt{\kappa^\mathcal{C}_2/N}$.  \emph{ii)} A second source of inaccuracy comes from the scatter of the value of the second cumulants across different conformers; a measure of this scatter is $\sigma_{\kappa_2}$, the standard deviation of the cluster distribution of second cumulants, and the corresponding contribution to the free-energy uncertainty is $\Delta F_2 = \frac{\sigma_{\kappa_2}}{2k_BT}$. \emph{iii)} Finally, the second energy cumulant within each conformer is affected by its own statistical error, $\Delta\kappa^\mathcal{C}_2$, whose contribution to the overall uncertainty is $\Delta F^\mathcal{C}_3 = \frac{\Delta\kappa^\mathcal{C}_2}{2k_BT}$. Combining these three contributions, the overall uncertainty on the free energy of the $\mathcal{C}$ conformer is estimated as:

\begin{equation}
\Delta F^\mathcal{C}=\sqrt{(\Delta F^\mathcal{C}_1)^2 + (\Delta
  F_2)^2 + (\Delta F^\mathcal{C}_3)^2},
\label{eq:error}
\end{equation}

\section{Results} \label{sec:results} 
\begin{figure}[h!]
  \centering
  \includegraphics[width=\hsize]{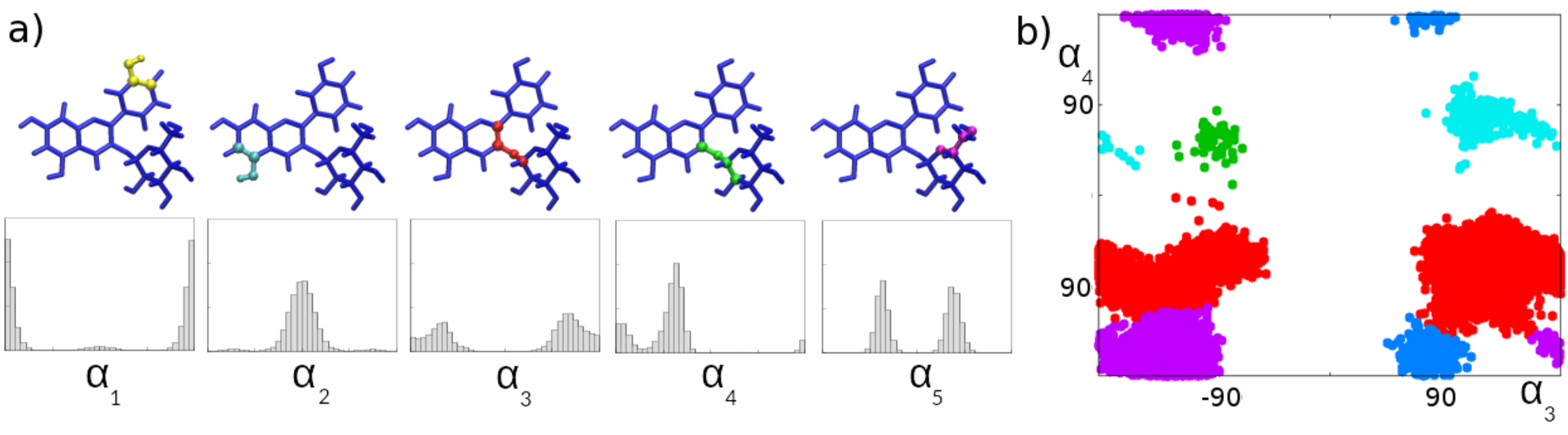}
  \caption{ Cyanin molecule in its neutral quinonoidal base state. (a) Upper row: the dihedrals highlighted in the different panels are those used to characterize the conformers. Lower row: probability distribution of the dihedrals depicted in the upper row, computed on a 2$\mu$s MM trajectory. (b): cluster representation of the conformational macrostates in the space of the $\alpha_3$ and $\alpha_4$ dihedrals; the different conformers found following Ref. \onlinecite{rodriguez2014clustering} are highlighted in different colors.  }
  \label{fig:uno} 
\end{figure}

As a case study, we tested our method on cyanin (cyanidin-3-glucoside, Figure \ref{fig:uno}). For this system we first ran a 2 $\mu$s MM trajectory using the Gromacs 4 MD package \cite{hess2008gromacs,van2005gromacs}. The cyanin FF (referred to in the following as ``$\mathrm{FF_{amb}}$'') was generated with the \emph{antechamber} tool \cite{wang2006automatic,wang2004development} and RESP charges were calculated with the RESP ESP charge Derive (R.E.D.)  program \cite{vanquelef2011red}.  The calculations were carried out in the NVT ensemble using a Nos\'e-Hoover thermostat in an orthorhombic cell with dimensions $19.22 \times 19.22 \times 16.53 ~\mathrm{\AA}^3$, filled with 177 water molecules. The slowest degrees of freedom for this molecule are dihedral rotations. Some of the relevant dihedrals are shown in Figure \ref{fig:uno}a together with their probability distribution evaluated along the MM trajectory. The distributions are peaked at a few maxima, signaling the presence of different conformers. The rotations of OH groups usually thermalize over AIMD time scales, with the exception of the $\alpha_1$ and $\alpha_2$ dihedrals depicted in Figure \ref{fig:uno}a.

In Figure \ref{fig:uno}b we display the joint probability distributions of two different dihedrals. As typical in molecules of this complexity, different dihedrals are highly correlated, making the identification of the relevant conformers a non trivial task. In order to achieve this goal in the space defined by the five dihedrals of Figure \ref{fig:uno}a, we adopted a newly developed clustering algorithm\cite{rodriguez2014clustering}. In this approach a cluster is defined as a peak in the joint probability density in the relevant space. The clusters identified by the algorithm in the subspace defined by the two dihedrals $\alpha_3$ and $\alpha_4$ are represented in different colors in Figure \ref{fig:uno}b. The density peaks are identified based on the observation that they have a larger density than neighboring configurations and that their minimum distance from points of higher density is anomalously large. In the case of water-solvated cyanin, we found 10 configurations satisfying this criterion in the space of the 5 dihedrals of Figure \ref{fig:uno}a. Once the density peaks have been thus identified, each of them is assumed to define a different conformer (cluster), and every other configuration is assigned to the cluster identified by the nearest density peak, following the procedure outlined in Ref. \onlinecite{rodriguez2014clustering}. The populations of these conformers according to the classical FF are defined as the number of configurations generated by the MMMD run and assigned to each of them by the clustering algorithm. Conformational populations cannot be evaluated in this way at the QM level of theory, because sufficiently long MD runs cannot be afforded in this case, and one has to resort therefore to a scheme based on free-energy corrections, following Eqs. \eqref{eq:Fdef} and \eqref{eq:sviluppo}.

A direct validation of this approach cannot be afforded either, because it would require a non-perturbative estimate of the QM populations, which is not feasible, as we have seen. For this reason, we decided to validate our perturbative approach using two different force fields, the first being the actual force field used in practice for our cluster analysis, $\mathrm{FF_{amb}}$, the second representing a ``classical proxy'' of the QM description of the system and dubbed ``$\mathrm{FF_{QM}}$''. We can then proceed to comparing the predictions of Eq. \eqref{eq:sviluppo} for different proxies ($\mathrm{FF^1_{QM}}, \mathrm{FF^2_{QM}}, \cdots$), designed so as to represent increasingly important differences between the QM and MM levels of theory, against accurate estimates based on long MD runs performed with the same FFs. The goal of this procedure is to identify suitable indicators enabling us to estimate the accuracy of the perturbative approach when a non-perturbative evaluation of the relative populations (or free-energy differences) is not feasible.

We stress that in all cases molecular configurations are classified in terms of the clusters identified from the trajectory generated through the \emph{bona fide} MM force field, $\mathrm{FF_{amb}}$. ``QM'' conformational populations are estimated by assigning each configuration of the $\mathrm{FF_{QM}}$ trajectory to the closest cluster as defined by clustering the $\mathrm{FF_{amb}}$ trajectory. In this way $\mathrm{FF_{QM}}$ populations can be estimated as the number of geometries belonging to each conformer, and the corresponding free energies evaluated from Eq. \eqref{eq:Fdef}. The typical auto-correlation time of the energy within each conformer is of the order of 3-4 ps. In order to sample QM-MM energy differences, 2000 configurations per conformer were thus selected along the MM trajectory with a time lag of 10 ps. Second energy cumulants were also evaluated over this same set of configurations, once water molecules had been stripped off from them, thus permitting to considerably reduce statistical fluctuations, without compromising our aim to utilize second cumulants to estimate the accuracy of first-order free-energy differences.

\begin{figure}[h!]
 \centering
 \includegraphics[width=15cm]{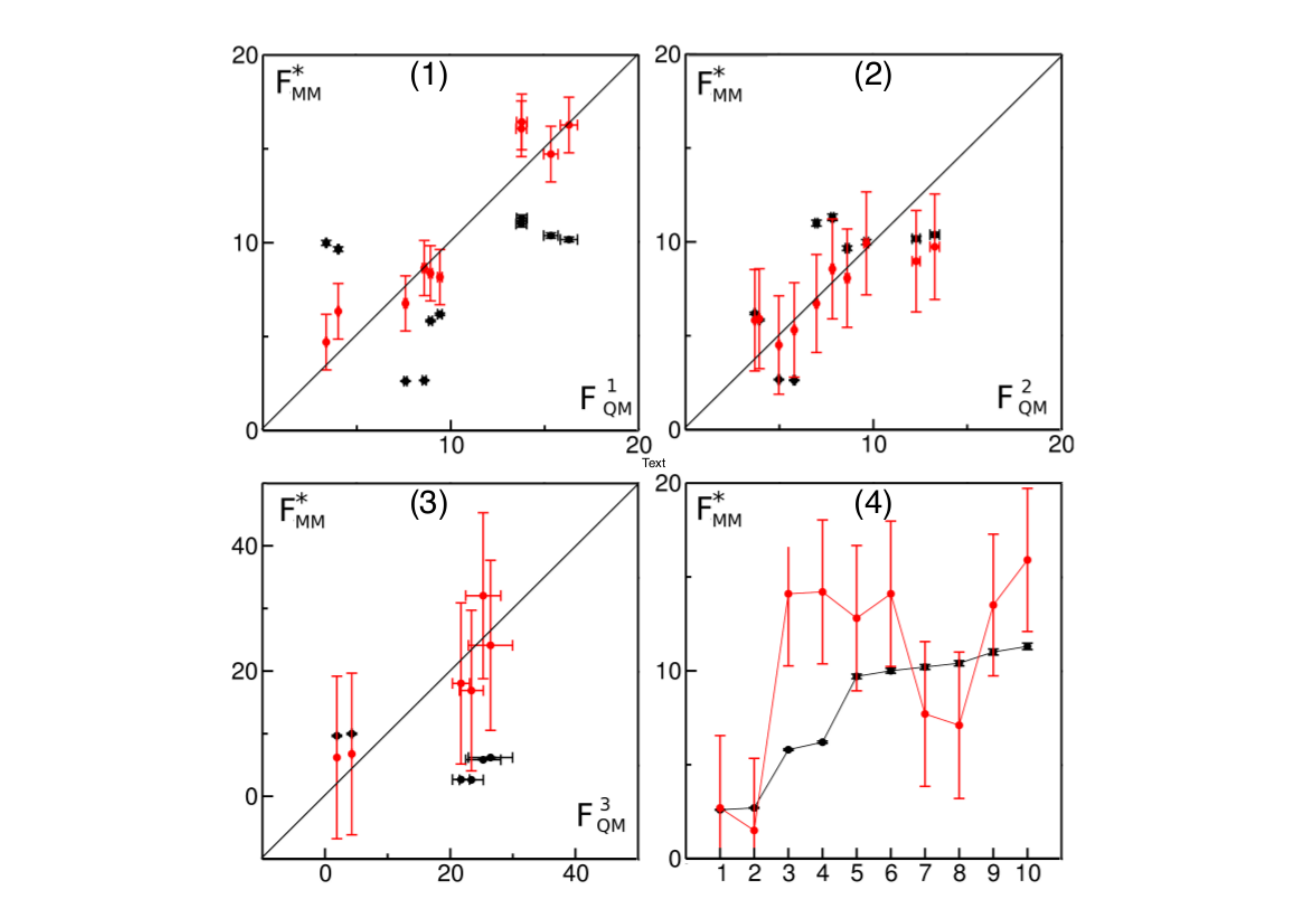}
 \caption{ 
Panels 1-3. Horizontal axis: $\mathrm{FF_{QM}}$ is the ``quantum" free energy of different molecular conformers computed with different ``classical proxy" force fields ($\mathrm{FF^1_{QM}}$, $\mathrm{FF^2_{QM}}$, and $\mathrm{FF^3_{QM}}$ in panels 1-3, respectively: see text). Vertical axis: $\mathrm{FF^*_{QM}}$ is the free energy computed for each conformer using Eq. \eqref{eq:sviluppo}. Black symbols indicate purely ``classical" results obtained with the $\mathrm{FF_{amb}}$  force field and neglecting first order corrections in Eq. \eqref{eq:sviluppo}. Red symbols indicate results including first order corrections. Error bars are estimated from Eq. \eqref{eq:error}. Note the change of scale between panels 1-2 and 3. Panel 4: comparison of the classical and QM free energies (the latter computed at the DFT level using the perturbative scheme introduced in the present paper) of different molecular conformers of the cyanin quinoidal base. Units are kJ/mol throughout.
}
 \label{fig:due}
\end{figure}

In Figure \ref{fig:due} (a-c) we show the correlation plots between
$\mathrm{FF_{QM}}$ and $\mathrm{FF_{amb}}$ free energies for three
different QM proxy FFs, dubbed as $\mathrm{FF^1_{QM}}$,
$\mathrm{FF^2_{QM}}$, and $\mathrm{FF^3_{QM}}$, and designed so as to 
differ increasingly from the \emph{bona fide} MM FF, $\mathrm{FF_{amb}}$. In each one of these figures is reported the $\mathrm{FF_{QM}}$ free energy (x axis) correlated with the $\mathrm{FF_{amb}}$ free energy (black) and with the proxy QM free energy (red) as calculated from first-order thermodynamical perturbation theory from $\mathrm{FF_{amb}}$. The errors on the linear corrections are estimated from the second energy cumulants, as explained at the end of the previous section. One clearly sees that the magnitude of the estimated errors is a faithful indicator of the quality of the first-order approximation: free energies calculated with Eq. \eqref{eq:sviluppo} from $\mathrm{FF_{amb}}$ results can always be trusted within the estimated error. By studying \emph{a   posteriori} the free energy landscapes generated by the $\mathrm{FF^1_{QM}}$, $\mathrm{FF^2_{QM}}$, and $\mathrm{FF^3_{QM}}$ force fields, we notice how the corresponding configurational spaces become increasingly different from that of $\mathrm{FF_{amb}}$. In the extreme case of $\mathrm{FF^3_{QM}}$ the two systems are so different that the 10 most populated conformers of FF do not correspond anymore to the most populated conformers of $\mathrm{FF^3_{QM}}$ (this is the reason why in Figure \ref{fig:due}-c only six conformers are shown).

Finally, we applied our method to evaluate the QM free energies obtained at the DFT level. DFT calculations were performed with the \textsc{Quantum ESPRESSO} package version 5.0,\cite{QE-2009} with the same cell and number of solvent molecules as used in the MM simulations illustrated above, following the same procedure as with the proxy QM FFs.  DFT estimates of the various free energies are shown in Figure \ref{fig:due}-DFT. The final value of the estimated error is very similar for the different clusters and is on average 3.8 kJ/mol. The magnitude of the error can be further reduced by increasing the size of the configuration sample and/or by designing a classical FF that more closely mimics the DFT inter-atomic interactions, so as to improve the quality of predictions based on first-order perturbation theory.  We stress that the estimate of the error is also a function of the number of conformers we consider. In the present case, for instance, if we focus on the free energies of the conformers differing only for the orientation of the sugar (1, 2, 3, 4, 8, and 9) the error estimates lowers to $\approx 2.5 \mathrm{kJ/mol}$, showing a very good performance of the FF. Focusing on the pairs of conformers 1-2, 4-5, and 6-7, which differ for the orientation of $\alpha_1$ or $\alpha_2$ dihedrals (Figure \ref{fig:uno} a), instead, the error estimate raises to $3.5~ \mathrm{kJ/mol}$.

In order to get further insight into the free-energy differences predicted by our methodology, AIMD simulations were started from geometries belonging to conformers 1, 2, and 3. The latter
trajectory was observed to move spontaneously towards a more stable conformer (1 or 2) after few ps of dynamics. This result shows that conformer 3 is unlikely to be stable at the QM level of theory, consistently with large free-energy differences between it and conformers 1-2 ($> 10 \mathrm{kJ/mol}$) and with a very low barrier in-between, which, if existing, is easily crossed at room temperature.

\section{Conclusions}
In this paper we have introduced a new method to sample and characterize the conformational space of complex molecular species in solution, using first-order thermodynamical perturbation theory to estimate quantum-mechanical corrections to classical molecular-dynamics results, and second-order perturbation theory to estimate the ensuing accuracy.

First-order perturbation theory has been widely used in the past to evaluate quantum-mechanical corrections to free-energy differences, as estimated from classical molecular dynamics, particularly in the study of chemical reactions and solvation free energies\cite{wesolowski1994ab, brandsdal2003free, hu2008free,   rosta2006towards,woods2008efficient}. The scope of this methodology when applied to complex molecular systems is limited by the ability of the low level of theory (MM in our case) to describe the zero-th order conformational landscape with sufficient accuracy. Whilst this description can be systematically improved, at least in principle, by improving the MM force field, no reliable criteria have been available so far to evaluate the quality of the first-order correction. One of the main steps forward made in our work is the identification of such a quantitative criterion, based on a careful analysis of an approximate evaluation of the second-order correction. Our analysis also indicates that first-order corrections need not be small in order to be accurate: the accuracy only depends on the ability of the low level of theory to correctly describe the topography of the conformers, which in turn can be assessed by the accuracy criterion mentioned above.

A second important element of this work is a novel procedure for identifying the conformers on which perturbation theory is applied. A conformer is defined as a peak in the probability density in the possibly high-dimensional space spanned by the internal coordinates of the solute. These peaks are identified by a novel clustering algorithm\cite{rodriguez2014clustering}. Since the analysis is performed at finite temperature and in the presence of a solvent, the conformers do not necessarily coincide with the minima of the potential energy surface.  A correct definition of the conformers is a crucial ingredient in the procedure since, as we already mentioned, an error in the description of the topography of the system dramatically impacts the accuracy.

Of course, the problem still remains that the perturbative evaluation of free-energy differences is hindered by statistical fluctuations, which crucially depend on the system size. In the present case size-extensive fluctuations due to an explicit account of the solvent can be significantly reduced by
relying on a QM/MM approach whereby the extensive contributions of the solvent cancel exactly when computing the difference between QM/MM and pure-MM energies. 

In short, we believe that our approach offers a route to probe the free energy landscape of highly mobile molecular species with a QM level of description {\it by keeping the statistical accuracy under control}. The availability of a reliable estimate of the error naturally opens the way to systematically improve the free-energy estimates by fine-tuning the low level of theory.

\bibliography{freeEn}
  
\end{document}